\documentclass[aps,prd,twocolumn,showpacs,amsmath,amssymb,nofootinbib,eqsecnum, preprintnumbers]{revtex4-1}
\usepackage{graphicx}
\usepackage{epstopdf}
\usepackage{epsfig}

\def\appendix{{\newpage\section*{Appendix}}\let\appendix\section%
        {\setcounter{section}{0}
        \gdef\thesection{\Alph{section}}}\section}

\newcommand{\be}{\begin{equation}}
\newcommand{\ee}{\end{equation}}
\newcommand{\bear}{\begin{eqnarray}}
\newcommand{\eear}{\end{eqnarray}}
\newcommand{\ba}{\begin{array}}
\newcommand{\ea}{\end{array}}

\begin{document}

\title{The Extended Thermodynamic Properties of\\ a topological
Taub-NUT/Bolt-AdS spaces}
\author{Chong Oh Lee}
\email{cohlee@kunsan.ac.kr}
\affiliation{Department of Physics,
Kunsan National University, Kunsan 573-701, Korea}

\begin{abstract}
We consider higher dimensional topological Taub-NUT/Bolt-AdS solutions
where a cosmological constant is treated as a pressure.
The thermodynamic quantities of these solutions are explicitly calculated.
Furthermore, we find these thermodynamic quantities satisfy the Clapeyron equation.
In particular, a new thermodynamically stable region for the NUT
solution is found by studying the Gibbs free energy.
Intriguingly,
we also find that like the AdS black hole case, the $G-T$ diagram of
the Bolt solution has two branches which are joined at a minimum temperature.
The Bolt solution with the large radius, at the lower branch, becomes stable beyond
a certain temperature
while the Bolt solution with the small radius, at the upper branch, is always unstable.
\end{abstract}
\maketitle
\section{Introduction}
It has been found that the area of the event horizon
of a black hole is proportional to its physical entropy
in search for similarities between black hole physics and thermodynamics \cite{Bekenstein:1973ur}.
It has been successively suggested that
by using the thermodynamic relationship between the
thermal energy, temperature, and entropy,
the first law of black hole thermodynamics can be expressed in the similar forms to the first
law of standard thermodynamics \cite{Bardeen:1973gs}.
It has been also found that there is a phase transition in the
Schwarzschild-AdS black hole through investigations of complete analogy between black hole system
and standard thermodynamic system \cite{Hawking:1982dh}. Since then,
it has been studied for the phase transitions
and critical phenomena in a variety of black hole solutions \cite{Lousto:1994jd}-\cite{Myung:2012xc}
and extensively investigated in various thermodynamic issues of black hole in higher dimensional AdS space
\cite{Chamblin:1999tk}-\cite{Banerjee:2011au}.

It has been recently suggested that
by considering $(d+1)$-dimensional AdS black holes,
the thermodynamic pressure $p$ is given by
\bear\label{pressure}
p=-\frac{1}{8\pi}\Lambda=\frac{u(2u+1)}{8\pi l^2},
\eear
in units where $G=c=\hbar=k_B=1$, and
$u$ is of the form $d+1 = 2u+2$ with  a positive integer $u$.
Several series of relevant investigations have been performed
\cite{Kastor:2009wy}-\cite{Belhaj:2015hha}.

Recently, it has been shown that the thermodynamic volume in the
Taub-NUT-AdS case can be negative.
This negative
thermodynamic volume may be interpreted in that the environment
(universe) applies work to the system (Taub-NUT-AdS black hole) in
the process of the Taub-NUT-AdS black hole formation, while the
positive thermodynamic volume may be interpreted as applying the work
on the environment (universe) by the system (the whole black hole)
considering the process of forming the black hole
\cite{Johnson:2014yja}. They also have found that there is the first
order phase transition from Taub-NUT-AdS to Taub-Bolt-AdS
with considering the phase structure of these black holes
\cite{Johnson:2014pwa}. This issue has been studied for the Kerr-Bolt-AdS
case \cite{MacDonald:2014zaa} and investigated extensively in higher
dimensional NUT/Bolt case. In these higher dimensional cases, it has been
particularly found that the Taub-NUT-AdS solution has a
thermodynamically stable range as a function of the temperature for
any odd $u$ and there is the transition from Taub-NUT-AdS to
Taub-Bolt-AdS for all odd $u$ only \cite{Lee:2014tma}.

Furthermore, it has been shown that this higher dimensional NUT/Bolt case with a
discrete parameter $k$ can be generalized \cite{Mann:2005ra}. In the
context of the extended thermodynamics, the thermodynamic properties
of the case $k = 1$ has been studied \cite{Lee:2014tma} only. Thus, it would be
interesting to be a similar discussion of the
generalizations in the $k=0, -1$ topological solutions. More
intriguingly, their thermodynamic phase structure would be investigated
through exploring the behaviour of the Gibbs free energy since
understanding its behaviour is essential for uncovering possible
thermodynamic phase transitions. In this paper, we address these
questions.

The paper is organized as follows: in the next section we
investigate thermodynamic properties in topological Taub-NUT/Bolt-AdS spaces for
any $u$. We explicitly obtain the general forms of thermodynamic quantities such as the entropy,
the enthalpy, the specific heat, the temperature, the thermodynamic volume,
the Gibbs free energy, and the latent heat. In particular, by introducing
the Gibbs free energy we discuss their phase structure and their
instability. In the last section we give our conclusion.

\section{Topological Taub-NUT/Bolt-AdS Spaces}
We consider topological Taub-NUT/Bolt-AdS metric in higher dimensional spacetime
and the general solution in the Euclidean section is given by
\cite{Chamblin:1998pz}-\cite{Astefanesei:2004kn}
(for the generalized versions of the issue, see e.g., \cite{Mann:2005ra})
\bear
\textstyle ds^2&=&\textstyle f(r)\Big\{dt_{E}+4N\sum_{i=1}^{u}f_k^2(\frac{\theta_i}{2})d\phi_i\Big\}^2
+\frac{dr^2}{f(r)}
\nonumber\\&&+\textstyle(r^2-N^2)\sum_{i=1}^{u}
\Big\{d\theta_i^2+f_k^2(\theta_i)d\phi_i^2\Big\},
\eear
where $N$ represents a NUT charge for the Euclidean section,
and the metric function $f(r)$ is found to be
\bear
\textstyle f(r)&=& \textstyle\frac{r}{(r^2-N^2)^u}\int^{r}
\Big\{\frac{(a^2-N^2)^u}{a^2}k+\frac{(2u+1)(a^2-N^2)^{u+1}}{l^2 a^2}\Big\}da\nonumber\\
&&~~~~~~\textstyle-\frac{2mr}{(r^2-N^2)^u}.
\eear
with a cosmological
parameter $l$ and a geometric mass $m$.
The discrete parameter $k$ takes the values 1, 0, $-1$ and defines the form of the function $f_k(\theta_i)$
\begin{eqnarray}  \label{f}
 f_{k}(\theta_i )=\left\{
\begin{array}{ll}
\sin \theta_i , & \mathrm{for}\ \ k=1 \\
\theta_i , & \mathrm{for}\ \ k=0 \\
\sinh \theta_i , & \mathrm{for}\ \ k=-1,%
\end{array}%
\right.
\end{eqnarray}%
and the space ${\cal M}^2$ corresponds to a two dimensional sphere for $k=1$, plane for $k=0$,
and pseudohyperboloid for $k=-1$, respectively.
For $k=1$, the NUT solution occurs when solving $f(r)|_{r=N}$=0. The inverse of the temperature $\beta$
is obtained by requiring regularity in the Euclidean time
$t_{E}$ and radial coordinate $r$
\cite{Chamblin:1998pz}-\cite{Astefanesei:2004kn}
\bear\label{HT0}
\textstyle\beta=\left.\frac{4\pi}{f'(r)}\right|_{r=N}=\frac{4(u+1)\pi}{\sigma} N,
\eear
where $\sigma$ is a positive integer and $\beta$  is the period of $t_{E}$. The $\sigma$ appears
since the period cannot be bigger than $4(u+1)\pi N$, so that the Misner-string singularities vanish.
However, when $\sigma$ has an integer value, the period can smaller than $4(u+1)\pi N$.
This property is not guaranteed for $k=0$ and $k=-1$ since there is no Misner-string and no periodicity of time $t$.
However, it was shown that this holds for the $k=0,-1$ cases
through checking the self-consistency of the thermodynamic relations \cite{Astefanesei:2004kn}.

Let us first consider the NUT solution ($r=N$). Then after taking $\sigma=1$ for convenience without loss of generality,
we obtain the following formula for the temperature:
\bear\label{T1}
\textstyle\frac{1}{\beta}=T=\frac{k}{4(u+1)\pi N}
\eear
where $k$ has 1 or 0 since there are no hyperbolic NUT solutions.
Thus, for the NUT solution we consider the cases $k=1,0$ only.

Using counter term subtraction method, we get the regularized action
\cite{Chamblin:1998pz}-\cite{Astefanesei:2004kn}
\bear\label{ac0}
\textstyle I_{\rm NUT}=\frac{(4\pi)^{u}N^{2u-1}
(2uN^2-kl^2)}{16\pi^{\frac{3}{2}}l^2}
\Gamma(\frac{1}{2}-u)\Gamma(u+1)\beta,
\eear
where the gamma function $\Gamma(t)$ is defined as
$\Gamma(t)=\int_{0}^{\infty}x^{t-1}e^{-x}dx$.

Employing the Gibbs-Duhem relation $S=\beta M - I$ and substituting in (\ref{pressure}),
the entropy is found to be
\bear\label{s0}
\textstyle S_{\rm NUT}
&=&\textstyle\frac{(4\pi)^{u}N^{2u-1}\Big\{16\pi N^2p-(2u-1)k\Big\}}{16\pi^{\frac{3}{2}}}\nonumber\\
&&\textstyle\times\Gamma(\frac{1}{2}-u)\Gamma(u+1)\beta.
\eear
Here $M$ is the conserved mass $M=u(4\pi)^{u-1}m$,
and this will be identified with enthalpy $H$ ($M\equiv H= U+pV$) \cite{Kastor:2009wy}, which leads to
\bear
\textstyle H_{\rm NUT}&=&\textstyle u(4\pi)^{u-1}\Gamma(\frac{3}{2}-u)
\Gamma\left(u+1\right)
\Big\{\frac{N^{2u-1}}{\sqrt{\pi}(2u-1)}k\nonumber\\
&-&\textstyle\frac{16\sqrt{\pi}(u+1)N^{2u+1}}{u(2u-1)(2u+1)}p\Big\}.
\eear

One the other hand, by thermal relation $C=-\beta \partial_{\beta}S$, the specific heat is given as
\bear
\textstyle C_{\rm NUT}&=&\textstyle\frac{2\pi (4u+1)(u+1)^2S_{\rm NUT}T^2}{\pi(2u-1)(u+1)^2T^2-kp},
\eear
which is negative for $p>\pi(2u-1)(u+1)^2T^2/k$ while is positive $p<\pi(2u-1)(u+1)^2T^2/k$
and diverges at $p=\pi(2u-1)(u+1)^2T^2/k$.

The thermodynamic volume is also obtained as
\bear\label{NUTVol}
\textstyle V_{\rm NUT}&=&\textstyle-\frac{u(4\pi)^uN^{2u+1}}{2\sqrt{\pi}}\Gamma(-\frac{1}{2}-u)\Gamma(u).
\eear

Then from the above thermodynamic quantities,
the generalized Smarr formula due to dimensional scaling arguments for
any value of $k$ is given as
\bear\label{smarr}
\frac{1}{2}H-\frac{u}{2u-1}TS+\frac{1}{2u-1}pV=0,
\eear
which is precisely matched with that of static $d$-dimensional black holes with negative cosmological constant
\cite{Caldarelli:1999xj,Kastor:2009wy,Cvetic:2010jb}.

Using an thermal relation $U=H-pV$, the internal energy of Taub-NUT is obtained as
\bear
\textstyle U_{\rm NUT}&=&\textstyle\frac{u(2u+1)}{8\pi}\Big\{(2\sqrt{\pi}N)^{2u-1}k
-\frac{(2u+1)(2\sqrt{\pi}N)^{2u-1}N^2}{l^2}\Big\}\nonumber\\
&&~~~~~~\textstyle\times\Gamma(-\frac{1}{2}-u)\Gamma(u+1).
\eear

\begin{figure}[!htbp]
\begin{center}
{\includegraphics[width=8cm]{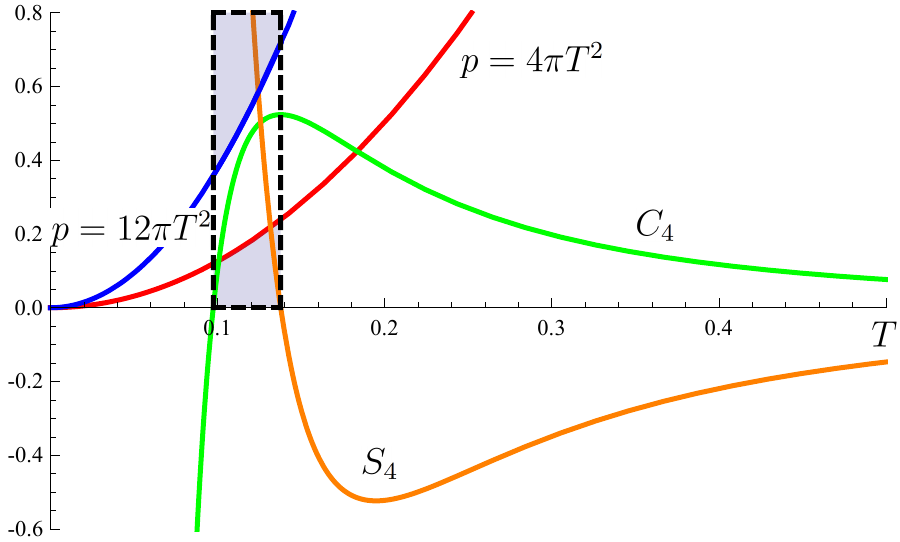}}
\end{center}
\vspace{-0.6cm}
\caption{{\footnotesize Plot of the entropy $S_4$ (yellow solid curve),
specific heat $C_4$ (green solid curve) and pressure $p$ (red solid curve for $p=4\pi T^2$, and
blue solid curve for $p=12\pi T^2$, respectively)
as a function of the temperature $T$ in four dimensions for $k=1$.}}
\label{figI}
\end{figure}
Finally, employing the Legendre transform of enthalpy $G=H-TS$, the Gibbs free energy is given as
\bear\label{Gibbs1}
\textstyle G _{\rm NUT}&=&\textstyle\frac{k^{2u}}{(2\sqrt{\pi})^{2u+3}}
\frac{\pi(2u+1)(u+1)^2T^2-kp}{\big\{(u+1)T\big\}^{2u+1}}\nonumber\\
&&~~~~~~~~~\textstyle\times\Gamma(-\frac{1}{2}-u)\Gamma(u+1).
\eear
Requiring both the entropy and the specific heat for $k=1$ are positive,
we get the following thermally stable range of $T$ for the NUT solution
\bear
\textstyle\sqrt{\frac{p}{\pi(2u-1)(u+1)^2}}< T <\sqrt{\frac{p}{\pi u(2u-1)(u+1)}}.
\eear
For example, the NUT solution in four dimensions is thermally stable in the region
inside dashed box in Fig. 1.
since both the entropy and the specific heat are positive when $\sqrt{\frac{p}{4\pi}}<T<\sqrt{\frac{p}{2\pi}}$.
However when the Gibbs free energy is introduced, the NUT solution in some areas of this region
is still unstable since the Gibbs free energy (\ref{Gibbs1}) is positive for
$p<4\pi T^2$ or $p>12\pi T^2$
(as you see in Fig. 1, the NUT solution in the shaded areas
is thermally unstable), that is to say that the Taub-NUT-AdS system
evaporates to a stable cold remnant (the pure AdS spacetime)
whereas the Taub-NUT-AdS system is a more thermally
stable configuration than the pure AdS spacetime since the Gibbs free energy is negative
for $4\pi T^2<p<12\pi T^2$.
For any $u$, the NUT solution is thermally unstable since the Gibbs free energy (\ref{Gibbs1}) is positive for
$p<\pi(2u-1)(u+1)^2 T^2$ or $p>\pi(2u+1)(u+1)^2 T^2$ while the NUT solution is thermally stable
since the Gibbs free energy is negative for $\pi(2u-1)(u+1)^2 T^2<p<\pi(2u+1)(u+1)^2 T^2$.

\begin{figure}[!htbp]
\begin{center}
{\includegraphics[width=8cm]{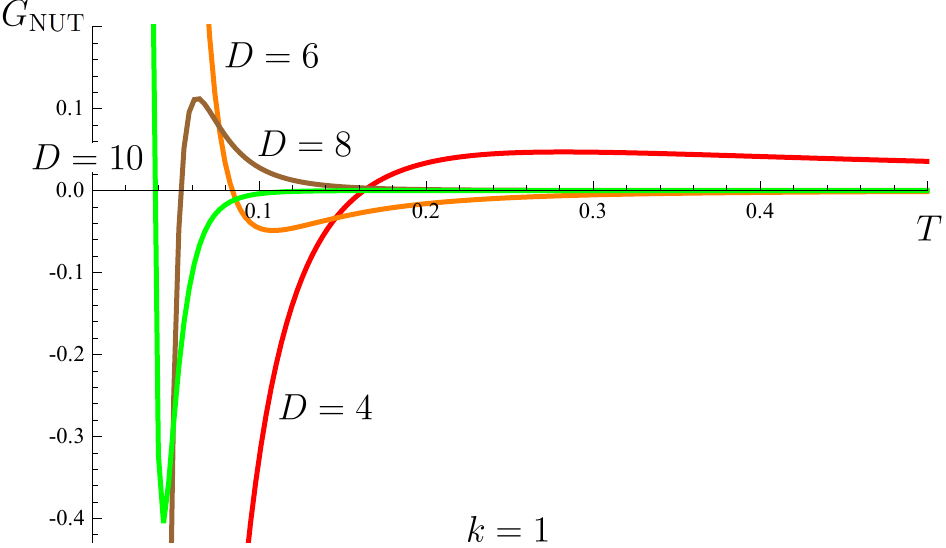}}
\end{center}
\vspace{-0.6cm}
\caption{{\footnotesize Plot of the $(2u+2)$-dimensional Gibbs free energy $G_{\rm
NUT}$ as a function of temperature $T$ for $k=1$ \& $p=1$ (red solid
curve for $u=1$, yellow solid curve for $u=2$, brown solid curve for
$u=3$, and green solid curve for $u=4$, respectively).}}
\label{figII}
\end{figure}

For $k=1$, as shown in Fig 2., when the temperature $T$ is bigger than a certain temperature,
the Gibbs free energy is positive for odd $u$ while the Gibbs free energy is negative
for even $u$. In this range of the temperature $T$
the phase transition from the Taub-NUT-AdS spaces to the pure AdS spacetime occurs for odd $u$
whereas that of the opposite direction occurs for even $u$.
It is quite natural that this thermodynamic behavior of the Taub-NUT-AdS spaces
changes due to odd $u$ or even $u$ since the sign of the gamma function in the Gibbs free energy (\ref{Gibbs1})
alternates due to odd $u$ or even $u$.

Remarkably, this topological Taub-NUT-AdS solution has the Hawking-Page transition
between the Taub-NUT-AdS spaces and the pure AdS spacetime, which occurs on the line
$\tilde{p}_{\rm coex}={(2u+1)(u+1)^2 \pi T^2}/{k}$, or
$S_{\rm NUT}=\frac{\Gamma(\frac{1}{2}-u)\Gamma(u+2)}{2\sqrt{\pi}}
\left\{\frac{k}{2\sqrt{\pi}(u+1)}\right\}^{2u}
\left(\frac{1}{T}\right)^{2u}$.

As shown in Fig 3.,
there is the NUT solution phase for $p<\tilde{p}_{\rm coex}$ at any fixed temperature
while there is AdS phase for $p>\tilde{p}_{\rm coex}$ and
the two states exist together for $p=\tilde{p}_{\rm coex}$.
The coexistence lines of the topological Taub-NUT-AdS solution phases
and the divergent lines of the topological Taub-NUT-AdS solution specific heat
move more to the left as the dimension of spacetime increases.

\begin{figure}[!htbp]
\begin{center}
{\includegraphics[width=8cm]{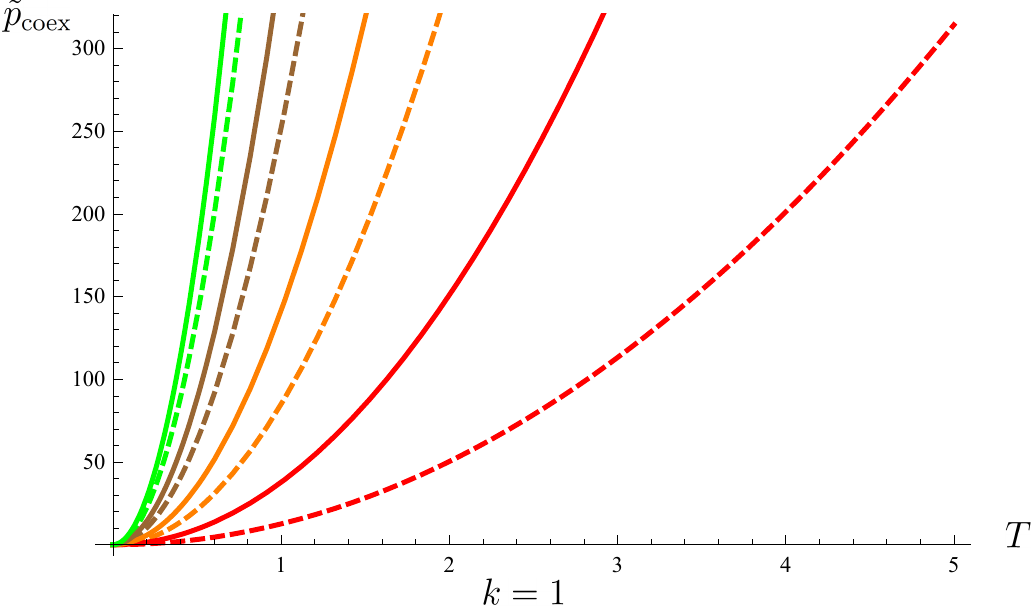}}
\end{center}
\vspace{-0.6cm}
\caption{{\footnotesize For $k=1$, the coexistence lines of the
Hawking-Page phase transition are solid curves and the specific heat diverges on dashed curves
(red solid/dashed curve for $u=1$, orange
solid/dashed curve for $u=2$, brown solid/dashed curve for $u=3$, and green solid/dashed
curve for $u=4$, respectively).
(For interpretation of the references to color in this figure legend,
the reader is referred to the web version of this article.)}} \label{figIII}
\end{figure}

The shift of entropy  occurs on the coexistence lines of the topological Taub-NUT-AdS solution phases,
and so the displacement of entropy is given by
\bear\label{Ds}
\textstyle\Delta S=\frac{\Gamma(\frac{1}{2}-u)\Gamma(u+2)}{2\sqrt{\pi}}
\left\{\frac{k}{2\sqrt{\pi}(u+1)}\right\}^{2u}
\left(\frac{1}{T}\right)^{2u},
\eear
which leads to the latent heat
\bear\label{LH1}
\textstyle L_{\rm NUT}=\frac{\Gamma(\frac{1}{2}-u)\Gamma(u+2)}{2\sqrt{\pi}}
\left\{\frac{k}{2\sqrt{\pi}(u+1)}\right\}^{2u}
\left(\frac{1}{T}\right)^{2u-1},
\eear
by an thermal relation $L=T\Delta S$, and
the latent heat in terms of $p$ is obtained as
\bear\label{LH2}
\textstyle L_{\rm NUT}=\frac{\big\{4(2u+1)\big\}^{u-\frac{1}{2}}}{\pi}\Gamma(\frac{1}{2}-u)\Gamma(u+2)
\left(\frac{1}{p}\right)^{u-\frac{1}{2}}.
\eear
Then at any given temperature $T$ for any $u$, the displacement of entropy and
the latent heat are zero in the case $k=0$
and for odd $u$ are negative in the case $k=1$.
The former is trivial since the temperature $T$ (\ref{T1}) is zero
as well as the Gibbs free energy $G_{\rm NUT}$ (\ref{Gibbs1}). However, the latter is
related with energy supplied by the process of forming the Taub-NUT-AdS system
from the pure AdS spacetime. It indicates
a net release of latent energy back into the environment
because of evaporating of Taub-NUT-AdS system.
The latent heat (\ref{LH1}), like the AdS black hole case \cite{Dolan:2014mra}, vanishes
as the temperature goes to infinity instead of a certain finite value (critical temperature)
and a second order phase transition cannot take place at a certain finite temperature.
The pressure $p$ vanishes for asymptotically flat spacetime and
the latent heat (\ref{LH2}) becomes infinity. This means that it is not possible to spontaneously form the black
hole from the Minkowski spacetime.

For AdS spacetime the Gibbs free energy is $G_{\rm AdS}=0$
and for the Taub-NUT-AdS solution
\bear
G_{\rm NUT}=U_{\rm NUT}-TS_{\rm NUT}+pV_{\rm NUT}.
\eear
Their differentials on the coexistence line are $dG_{\rm AdS}=0$
and $dG_{\rm NUT}=-S_{\rm NUT}\,dT+V_{\rm NUT}\,dp$, which lead to
\bear\label{dG0}
0=dG_{\rm AdS}-dG_{\rm NUT}=S_{\rm NUT}\,dT-V_{\rm NUT}\,dp
\eear
where the coexistence line with two states may be defined by $G_{\rm AdS}=G_{\rm NUT}$.
Hence, the insertion of (\ref{s0}) and (\ref{NUTVol}) into (\ref{dG0}) yields
\bear\label{Cle}
\textstyle \frac{dp}{dT}=\frac{S_{\rm NUT}}{V_{\rm NUT}}=\frac{2\pi(2u+1)(u+1)^2}{k}.
\eear
Furthermore, since the thermodynamic volume and the entropy become zero for AdS spacetime
we can write
\bear
\Delta V=V_{\rm NUT}~~~{\rm and}~~~
\Delta S=S_{\rm NUT},
\eear
and have the Clapeyron equation ${dp}/{dT}={\Delta S}/{\Delta V}$.

As checking the self-consistency of the thermodynamic relations,
we can reproduce the above result (\ref{Cle}) through
$\frac{dp}{dT}=\frac{2\pi(2u+1)(u+1)^2}{k}$,
which shows that the Clapeyron equation still holds for the NUT solution.

Let us consider the Bolt solution ($r=r_{\rm B}>N$).
Requiring $f(r)|_{r=r_{\rm B}>N}$ and $f'(r)|_{r=r_{\rm M}}=\frac{1}{N(u+1)}$, the Bolt solution
occurs. In Taub-Bolt-AdS metric, the inverse of the temperature, the action,
and the mass are respectively
\bear\label{inverT}
\textstyle \beta=\left.\frac{4\pi}{f'(r)}\right|_{r=r_{\rm B}}
=\frac{4\pi l^2 r_{\rm B}}{kl^2+(2u+1)(r_{\rm B}-N^2)},
\eear
\bear
\textstyle I_{\rm Bolt}&=&\textstyle\frac{(4\pi)^{u-1}}{4 l^2}\Big[\frac{(2u+1)(-1)^u N^{2u+2}}{r_{\rm B}}\nonumber\\
&&\textstyle +\sum_{i=0}^{u}
\Big(\begin{array}{l}
u\\
i
\end{array}\Big)
(-1)^i N^{2i} r_{\rm B}^{2u-2i}\Big\{\frac{l^2}{(2u-2i-1)r_{\rm B}}k\nonumber\\
&&\textstyle -\frac{(2u+1)(u-2i+1)r_{\rm B}}{(2u-2i+1)(u-i+1)}\Big\}\Big]\beta.
\eear
Here the Bolt radius $r_{\rm B}$ is
\bear\label{rB}
\textstyle r_{\rm B,\pm}=\frac{l^2\pm\sqrt{l^4+(2u+1)(2u+2)^2N^2[(2u+1)N^2-kl^2]}}
{(2u+1)(2u+2)N},
\eear
where $r_{\rm B, +}$ denote the large radius of the Bolt solution
and  $r_{\rm B, -}$ is the small radius of the Bolt solution.
The discriminant of the square root in the Bolt radius (\ref{rB}) is always
positive for the $k =0, -1$ cases but sometimes negative for $k=1$, and so
the former has no upper limit on $N$ while the latter has the maximum magnitude of the NUT charge
$N_{\rm max}$
\bear
\textstyle N\leq \frac{l}{\sqrt{2(u+1)(2u+1)[u+1+\sqrt{u(u+2)}]}}=N_{\rm max}.
\eear
Furthermore, from the Bolt radius (\ref{rB}) one can find in the cases $k=0,-1$ the large radius
$r_{\rm B, +}$ exists only for $N>0$
since the Bolt solution occurs for $r_{\rm B,\pm}>N$.

Using parallel way as in the case of the NUT solution,
the enthalpy $H_{\rm Bolt}$, the entropy $S_{\rm Bolt}$, and thermodynamics volume $V_{\rm Bolt}$
for the Bolt solution yields respectively
\bear
\textstyle H_{\rm Bolt}&=&\textstyle \frac{u(4 \pi)^{u-1}}{2}
\Big\{\sum_{i=0}^{u}
\Big(\begin{array}{l}
u\\
i
\end{array}\Big)
\frac{(-1)^i N^{2i}r_{\rm B}^{2u-2i-1}}{(2u-2i-1)}k\nonumber\\
&+&\textstyle \frac{8\pi}{u}\sum_{i=0}^{u+1}
\Big(\begin{array}{l}
u+1\\
~~~i
\end{array}
\Big)
\frac{(-1)^iN^{2i}r_{\rm B}^{2u-2i+1}}{(2u-2i+1)}p\Big\},
\eear
\bear
\textstyle S_{\rm Bolt}&=&\textstyle\frac{(4\pi)^{u-1}}{4}
\Big[\sum_{i=0}^{u}
\Big(\begin{array}{l}
u\\
i
\end{array}\Big)
\frac{(2u-1)(-1)^iN^{2i}r_{\rm B}^{2u-2i-1}}{2u-2i-1}k\nonumber\\
&&\textstyle +\Big\{
\sum_{i=0}^{u}
\Big(\begin{array}{l}
u\\
i
\end{array}\Big)
\frac{8\pi(2u^2+3u-2i+1)(-1)^{i}N^{2i}}{u(u-i+1)(2u-2i+1)}\nonumber\\
&&~~~~~\textstyle \times r_{\rm B}^{2u-2i+1}
+\frac{8\pi(2u-1)N^{2u+2}}{ur_{\rm B}}
\Big\}p\Big]\beta,
\eear
\bear
\textstyle V_{\rm Bolt}&=&\textstyle\frac{\pi^{u}(2r_{\rm B})^{u-1}}{2u+1}
\textstyle\Big\{-2(N^2-r_{\rm B}^2)
\Big(1-\frac{N^2}{r_{\rm B}^2}\Big)^u\nonumber\\
&&\textstyle +r_{\rm B}^2\Big(\frac{N}{r_{\rm B}}\Big)^{2u+1}
{\rm B} \Big(\frac{N^2}{r_{\rm B}^2},\frac{1}{2}-u,u+1\Big)\Big\},
\eear
where the incomplete beta function
${\rm B} \left(x,a,b\right)$ is
defined as
${\rm B}\left(x,a,b\right)=\int_{0}^{x} t^{a-1}(1-t)^{b-1}dt$.
Like the previous NUT case, these thermodynamic quantities satisfy
the generalized Smarr formula (\ref{smarr}).

The specific heat is given as
\bear\label{Cboltup}
\scriptstyle C_{\rm Bolt,\pm}&=&
\scriptstyle\frac{u(2u+1)^2p}{2\sqrt{{\cal A}}{\cal B_{\pm}}^2T^2}\left(\frac{1}{4 \pi}\right)^u
\scriptstyle\left(\frac{\pi uT}{kp}+\frac{\sqrt{{\cal A}}}{(u+1)(2u+1)kpT}\right)^{2u}\nonumber\\
&&\scriptstyle\times\Big[\frac{T^2}{p}\left(1-\frac{k^4(2u+1)^2p^2}{{\cal B_{\pm}}^2}\right)^u\nonumber\\
&&\times\scriptstyle\Big\{\sqrt{{\cal A}}
\scriptstyle\Big(k^4p^2-\pi k^3 u (u+1)^2 p T^2-2\pi^2 u^2 (u+1)^2 T^4\Big)\nonumber\\
&&~\scriptstyle+(u+1)(2u+1)\Big(-k^7p^3+2\pi k^6 u (u+1)^2 p^2 T^2\nonumber\\
&&~~~\scriptstyle+\pi^2k^3u^2(u+1)^2pT^4-2\pi^3u^3(u+1)^2T^6\Big)\Big\}\nonumber\\
&&\scriptstyle+\frac{k^3}{\pi u (2u-1)}\Big(\pi u(u+1){\cal B_{\pm}}T^2+k^4(2u+1) p^2\nonumber\\
&&~~~~~~~~~~~~~~~\scriptstyle-2\pi k^3 u(2u+1)(u+1)^2pT^2\Big)\nonumber\\
&&~~~\scriptstyle\times \Big(-kp+\pi u (u+1) (2u-1) T^2\Big)\nonumber\\
&&~~~\scriptstyle\times{}_1F_2 \left(\frac{1}{2}-u,-u;\,\frac{3}{2}-u;\,\frac{k^4(2u+1)^2p^2}{{\cal B_{\pm}}^2}\right)\Big],
\eear
where $C_{\rm Bolt,+}$ denotes the specific heat with the radius $r_{\rm B,+}$, and
 $C_{\rm Bolt,-}$ is the specific heat with the radius $r_{\rm B,-}$.
Here, ${\cal A}$ and ${\cal B_{\pm}}$ are
\bear
&&{\cal A}=(2u+1)^2(k^4p^2-2\pi k^3u(u+1)^2pT^2+\pi^2u^2(u+1)^2T^4,\nonumber\\
&&{\cal B_{\pm}}=\sqrt{{\cal A}}\pm\pi u (u+1) (2u+1) T^2,\nonumber
\eear
and the hypergeometric function ${}_1F_2(a,b;c;,z)$ is defined for $|z| < 1$ by the power series
\bear
\textstyle{}_1F_2(a,b;c;,z)=\sum_{n=0}^{\infty}\frac{(a)_n(b)_n}{(c)_n}\frac{z^n}{n!}.
\eear
For example, the four-dimensional specific heat is obtained as
\bear\label{Cbolt4}
\scriptstyle C_{\rm Bolt,\pm}
&=&\scriptstyle-\frac{\pi T^2}{4k^2p^2}\scriptstyle\pm
\frac{(k^8 p^4-6\pi k^7 p^3 T^2+2\pi^2 k^4 p^2 T^4
\scriptstyle-8\pi^3k^3pT^6+8\pi^4 T^8)}{16\pi^2 k^2 p^2 T^4 \sqrt{k^4p^2-8\pi k^3 p T^2+4\pi^2 T^4}},\nonumber\\
\eear
which is well matched with the result in \cite{Astefanesei:2004kn}.
Here, as far as we know,
even if including the Bolt solution without a cosmological constant treated as a pressure,
the specific heat (\ref{Cboltup}) is firstly shown in the analytical expressions.
In fact, since the specific heat of the Bolt solution for arbitrary dimensions
has a highly complicated high-order polynomial terms, the analytical expressions have not been obtained
even for the case $k = 1$. Furthermore for $k=1$
the four-dimensional specific heat $C_{\rm Bolt,\pm}$ (\ref{Cbolt4}) diverges
at $T=T_2$
\bear\label{Tcu1}
\textstyle T_2=\sqrt{\frac{p}{\pi}+\frac{\sqrt{3}p}{2\pi}},
\eear
and for $k=1$ the higher dimensional specific heat (\ref{Cboltup}) also diverges
at $T$
\bear\label{Tcu}
\textstyle T=\frac{\sqrt{p}}{\sqrt{\pi u}}\sqrt{1+\frac{\sqrt{u}\sqrt{u+2}}{u+1}}.
\eear
The internal energy of AdS-Taub-Bolt $U$ is given as
\bear
\textstyle U_{\rm Bolt}&=&\textstyle\frac{(4\pi)^{u-1}}{4l^2}uN^{2u-1}\Big\{(2u+1)N^2-kl^2\Big\}\nonumber\\
&&~~~\textstyle\times{\rm B} \Big(\frac{N^2}{r_{\rm B}^2},\frac{1}{2}-u,u+1\Big),
\eear
and the Gibbs free energy
\bear
\textstyle G_{\rm Bolt}&=&\textstyle u\Big(\frac{N}{r_{\rm B}}\Big)^{2u}
{\rm B} \Big(\frac{N^2}{r_{\rm B}^2},\frac{1}{2}-u,u+1\Big)k+\frac{16\pi N r_{\rm B}}{2u+1}
\Big\{\nonumber\\
&&\textstyle\Big(1-\frac{N^2}{r_{\rm B}^2}\Big)^{u+1}-\Big(\frac{N}{r_{\rm B}}\Big)^{2u+1}
{\rm B} \Big(\frac{N^2}{r_{\rm B}^2},\frac{1}{2}-u,u+1\Big)\Big\}p.\nonumber\\
\eear

\begin{figure}[!ht]
\begin{center}
{\includegraphics[width=8cm]{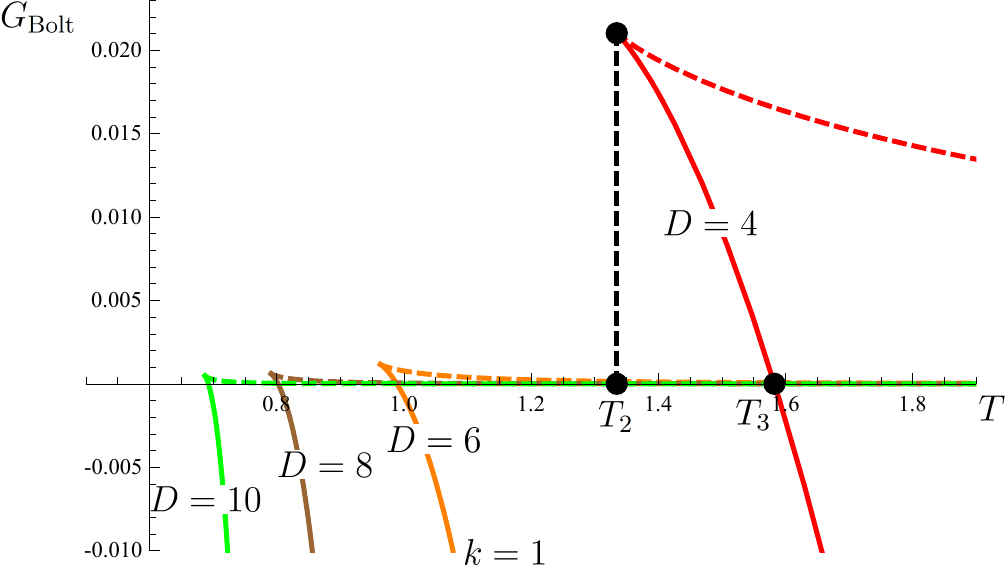}}
\end{center}
\vspace{-0.6cm}
\caption{{\footnotesize Plot of the $(2u+2)$-dimensional Gibbs free energy $G_{\rm Bolt}$ as a function of temperature $T$
for any $u$ (red solid/dahsed curve for $u=1$, yellow solid/dashed curve for $u=2$, brown solid/dashed curve
for $u=3$, and green solid/dashed curve for $u=4$, respectively)
for pressure $p=3$ and $k=1$.}}
\label{figIV}
\end{figure}

As shown in Fig 4., two branches (the red solid curve and the red dashed curve)
are joined at the temperature $T_2$. At this temperature two phases occur, that is to say that
the upper branch (red dashed curve) is the phase of the Taub-Bolt-AdS system with the small radius $r_{\rm B,-}$
and the lower branch (red solid curve) is the phase of the Taub-Bolt-AdS system with the large radius $r_{\rm B,+}$.
Like the AdS black hole case, at $T_2$ (\ref{Tcu1})
the four-dimensional specific heat of the Bolt solution also diverges and the higher dimensional cases diverge at $T$ (\ref{Tcu}).
Since the Gibbs free energy is positive, the Bolt solution for the upper branch is unstable
whereas since the lower branch includes the negative value of the Gibbs free energy,
this solution in such region becomes stable. Thus, when $G_{\rm Bolt}=0$,
like the Taub-NUT-AdS black hole case, there are the Hawking-Page transition
between the Taub-Bolt-AdS black hole
and the pure AdS spacetime, which occurs on the lines for the upper branch $r_{\rm B,+}$
\bear
\textstyle\bar{p}_{\rm coex,+}&=&\textstyle\Big[\frac{\pi^2 u (2u+1) (u+1)^2T^4}{k^5}\nonumber\\
&\times&\textstyle\frac{{\cal D}^{u+1}
B\left({\cal D},\frac{1}{2}-u,u+1\right)}{(1-{\cal D})^{u+1}/\sqrt{{\cal D}}+u {\cal D}^u
B\left({\cal D},\frac{1}{2}-u,u+1\right)}\Big]^{1/3},
\eear
with
\bear
\textstyle{\cal D}&=&\textstyle \frac{k^4p^2}{\left[\sqrt{k^4 p^2-2\pi u (u+1)^2 k^3 p T^2+\pi^2 u^2(1+u)^2T^4}
+\pi u (u+1) T^2\right]^{2}}\nonumber.
\eear

\begin{figure}[!ht]
\begin{center}
{\includegraphics[width=8cm]{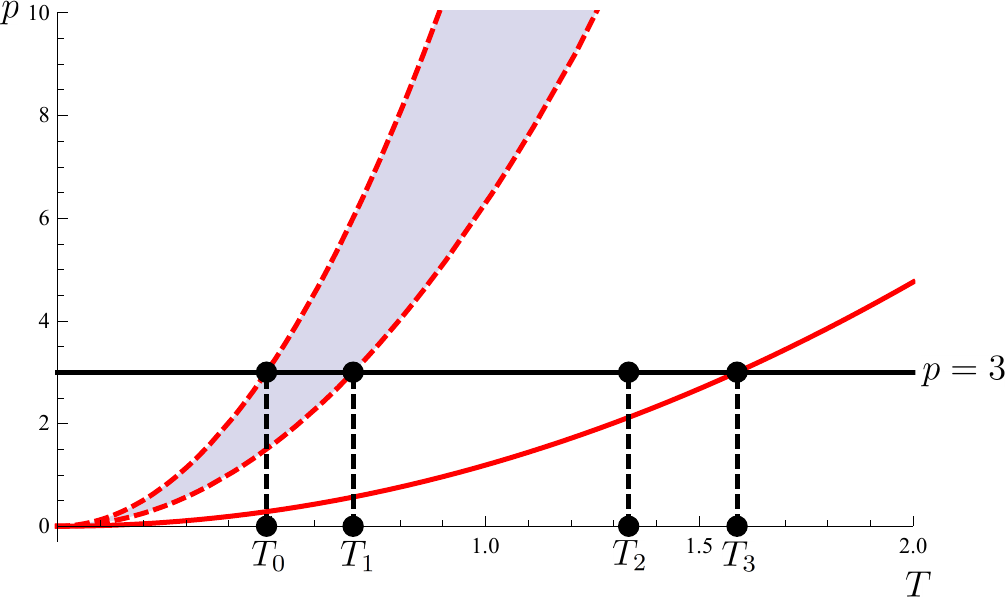}}
\end{center}
\vspace{-0.6cm}
\caption{{\footnotesize For the four-dimensional NUT solution and Bolt solution,
plot of $p$ as a function of $T$. Here $T_2$ and $T_3$ are the same values used in Fig 4.}}
\label{figV}
\end{figure}
Furthermore as shown in Fig 5., the thermodynamic instability of the Taub-NUT/Bolt-AdS system is classified
by the value of the temperature $T$. When $T>T_0$, the Taub-NUT-AdS system evaporates to
a stable cold remnant. When $T_0\leq T\leq T_1$, the Taub-NUT-AdS system
is a thermally stable configuration. When $T_1< T < T_2$, and then the Taub-NUT-AdS system evaporates again
since the four-dimensional entropy $S_4$ is negative. When $T_2\leq T < T_3$, the Taub-Bolt-AdS
system with two phases occurs at $T_2$ but still evaporates
since the Gibbs free energy of the Taub-Bolt-AdS system is positive.
Finally, when $T\geq T_3$, the Taub-Bolt-AdS system with the large radius $r_{\rm B,+}$ becomes stable
since the Gibbs free energy of the Taub-Bolt-AdS system is negative.
This thermodynamic nature of the Taub-NUT/Bolt-AdS system may hold for higher dimensional cases
since $p-T$ phase diagrams in higher dimension are similar shapes \cite{Lee:2014tma}.

From now on, considering the action difference of Taub-NUT-AdS and Taub-Bolt-AdS,
we investigate the instability of the Taub-NUT/Bolt-AdS system.
Their action difference, ${\cal I}_{D}$ is defined as $I_{\rm Bolt}-I_{\rm NUT}$
\bear\label{diffaction}
\textstyle {\cal I}_{D}&=&\textstyle\frac{(4\pi)^u}{8r_{\rm B}l^2}\Big[2N(N^2-r_{\rm B}^2)(u+1)
\left(1-\frac{N^2}{r_{\rm B}^2}\right)^u\nonumber\\
&&\textstyle +(2uN^2-kl^2)\Big\{(u+1)N^{2u}{\rm B}\Big(\frac{N^2}{r_{\rm B}^2},\frac{1}{2}-u,u+1\Big)\nonumber\\
&&~~~~~~~~~~~~~~\textstyle-\frac{2N^{2u}}{\sqrt{\pi}}\Gamma(\frac{1}{2}-u)\Gamma(2+u)\Big\}\Big].
\eear

\begin{figure}[!htbp]
\begin{center}
{\includegraphics[width=8cm]{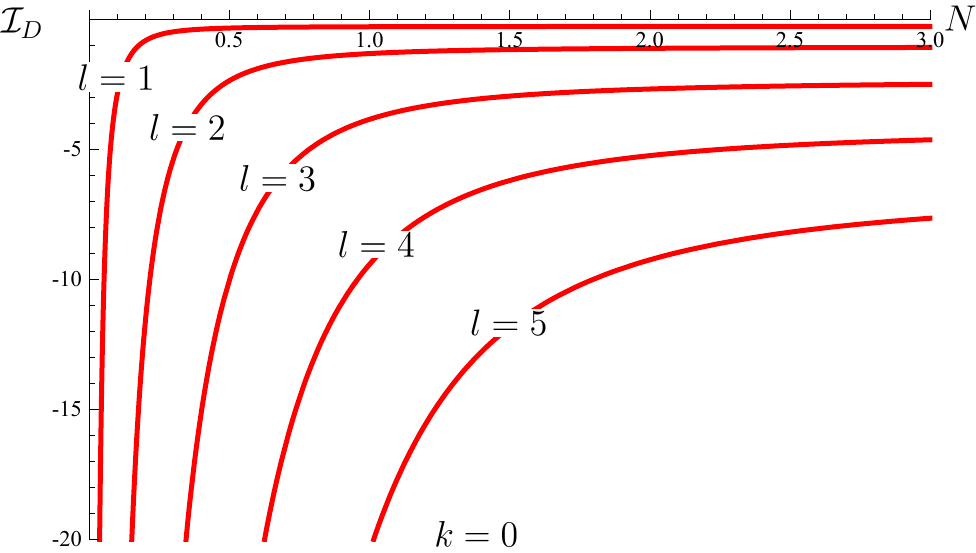}}
\end{center}
\vspace{-0.6cm}
\caption{{\footnotesize
Plot of the four-dimensional action difference ${\cal I}_4$ as a function of N for
cosmological parameter $l$ (from left to right) 1 to 5 for pressure $p = 3$ and $k=0$.
Solid lines are curves of ${\cal I}_4$ with large radius $r_{\rm B,+}$.}}
\label{figVI}
\end{figure}

Considering the four-dimensional case ($u=1$), for $k=1$
taking the cosmological parameter $l$ and the pressure $p$ as fixed parameters,
the action difference ${\cal I}_{D}$ becomes negative as $N$ increases.
This means that the Taub-Bolt-AdS system with $r_{\rm B,+}$
is a more thermally stable configuration than the Taub-NUT-AdS system
for ${\cal I}_{D}<0$, and so there is the first order phase transition from Taub-NUT-AdS system
to Taub-Bolt-AdS system with $r_{\rm B,+}$ at a critical NUT charge\cite{Johnson:2014pwa,Lee:2014tma}.
As shown in Fig 6., for the case $k=0$
Taub-Bolt-AdS system with $r_{\rm B,+}$ is stable only
since the lower branch is always negative whereas the upper branch is always positive.
It is shown
that the curves of ${\cal I}_4$ move more to the right as the pressure $p$ decreases
(the cosmological parameter $l$ grows up) since $p$ is inversely proportional
to $l$. For $u=2$ a similar result is obtained as with the $u=1$ case.
Furthermore, ${\cal I}-N$ diagrams in higher dimension are similar shapes \cite{Lee:2014tma}.
Thus, this thermodynamic instability of the Taub-NUT/Bolt-AdS system may hold
for higher dimensional cases, and is well matched with the result in \cite{Astefanesei:2004kn}.

\section{Conclusion}
We considered higher dimensional topological Taub-NUT/Bolt-AdS solutions
and in the context of the extended thermodynamics,
explicitly calculated their thermal quantities.
In particular, we gave the analytical expressions of the specific heat
for the $k = 1, 0, -1$ topological Bolt solutions.

By using these thermal quantities we also
investigated their phase structure.
By introducing Gibbs free energy, it was particularly found
that there is a new thermodynamically stable region of the NUT solution.

Furthermore, it was found that like the AdS black hole case, for $k=1$
Taub-Bolt-AdS system with two phases
occurs at a minimum temperature and at this temperature the specific heat diverges.

Recently using AdS/CFT correspondence,
thermodynamics and thermodynamic geometry have been covered for Schwarzschild-AdS black hole
in the extended phase space \cite{Zhang:2014uoa}.
It has been explored
through the extensive application of a RN-AdS black hole. They have shown that its thermodynamic behavior
is qualitatively the same as that in the Schwarzschild-AdS case \cite{Zhang:2015ova}.
It would be of interest to study the Taub-NUT/Bolt-AdS case in this context.


\end{document}